\documentclass[a4paper,12pt,oneside]{article}
\usepackage[english]{babel}
\usepackage{amsmath}
\usepackage{amssymb}
\usepackage{epsf}
\usepackage{graphicx}
\usepackage[usenames]{color}
\usepackage{tocenter}
\FromMargins[f]{15mm}{15mm}{20mm}{20mm}%

\renewcommand{\Im}{\mathop{\mathrm{Im}}\nolimits}
\renewcommand{\Re}{\mathop{\mathrm{Re}}\nolimits}

\title{Multiple Scattering of Waves in 3D Crystals (Natural or Photonic) Formed by Anisotropically
Scattering Centers}
\author{Baryshevsky V.G., Gurnevich E.A.\\
Research Institute for Nuclear Problems, Belarusian State
University,\\
11 Bobruiskaya Str., Minsk 220050, Belarus;\\
e-mail: bar$@$inp.bsu.by; genichgurn$@$gmail.com}
\date{}

\begin{document}
\maketitle

\begin{abstract}
This paper considers the refraction and diffraction of waves in
three-dimensional crystals formed by anisotropically scattering
centers. The partial wave expansion method is used to consider the
effect of multiple rescattering of waves by centers composing a
crystal.
The expression for the refractive index of a crystal is derived.
It is shown that instead of the diagonal elements of the
scattering matrix $\mathbf{T}$, appearing in the expression for
the refractive index of a chaotic medium, the derived expression
includes the diagonal elements of the reaction matrix
$\mathbf{K}$. This fact is taken into account in writing the
equations  describing the dynamical diffraction of waves in a
crystal.

The results can be of
interest for research into, e.g., diffraction of cold neutrons and
photons in crystals, nanocrystalline materials, as well as for the
description of  parametric and diffraction radiation in
electromagnetic crystals formed by anisotropically scattering
centers.

\end{abstract}

\maketitle

\section{Introduction}
For waves of different nature (photons, neutrons, etc.), the index
of refraction, $n$,  in the medium composed of chaotically
distributed identical centers can be expressed in terms of the
amplitude of elastic scattering by a single center as follows
\cite{lax,Goldberger,BVG95Nuclearoptics,landau77quant}:
\begin{equation}
  n^2 = 1 + \frac{4\pi\rho}{k^2}f(0).
  \label{eq:n2_3Dchaotic}
\end{equation}
Here $\rho$ is the density of scatterers (the number of scatters
per 1 cm$^3$ of matter) and $f(0)$ is the forward elastic
scattering amplitude, which is a complex number.

According to the optical theorem
\cite{Goldberger,BVG95Nuclearoptics,landau77quant}, the imaginary
part of the amplitude $f(0)$ of
 coherent elastic scattering at zero angle
 is related to the total cross section $\sigma$ of scattering
by the center  as
\begin{equation}
\label{2b}
 \Im f(0) = \frac{k\sigma}{4\pi},
\end{equation}
where $\sigma=\sigma_{el}+\sigma_r$ with $\sigma_{el}$ being the
elastic scattering cross section and $\sigma_r$ is the reaction
cross section. Thus, according to \eqref{eq:n2_3Dchaotic}, even in
the case of elastic scattering ($\sigma_r=0$) by individual
centers, the refractive index $n$ has an imaginary part describing
the attenuation of waves in the medium. Such attenuation
occurs because the waves scattered at nonzero angles acquire a
random phase due to chaotic distribution of scatterers.

When the waves are scattered in crystals, the situation is quite
different, because  the scatterers in this case are periodically
distributed in space \cite{BVG95Nuclearoptics,Batterman1964,lax2},
and the phases of rescattered waves are not random.
As a result, in the case of purely elastic scattering of waves by
the centers composing a  crystal, the refractive index should not
contain an imaginary part, i.e., it should be real. For example,
according to
\cite{BVG95Nuclearoptics,lax2,Baryshevsky1966}, for
isotropic scattering  of neutrons by crystal nuclei (the
scattering amplitude $f$ in this case is independent of the
scattering angle -- this is true for thermal and cold neutrons,
whose wavelength $\lambda$ is much larger than the nuclear radius
$R_n$) multiple rescattering of waves by the centers (nuclei),
composing the crystal, leads to the following expression for the
refractive index:
\begin{equation}
  n^2 = 1 + \frac{4\pi}{k^2\Omega_3}\frac{f}{1+ikf},
  \label{eq:n2_3Dcrystal}
\end{equation}
where $\Omega_3$ is the  unit cell volume of the crystal, i.e.,
\begin{equation}
  n^2 = 1 +
  \frac{4\pi}{k^2\Omega_3}\frac{f(1-ikf^*)}{|1+ikf|^2}=1+\frac{4\pi}{k^2\Omega_3}\frac{\texttt{Re}\,f}{|1+ikf|^2}.
  \label{3b}
\end{equation}
In deriving \eqref{3b} we used the optical theorem \eqref{2b} and
took into account that for isotropic elastic scattering
$\sigma=4\pi|f|^2$. As we can see from \eqref{3b}, multiple
rescattering of waves in crystals causes the refractive index to
become real.

Let us recall that \eqref{eq:n2_3Dcrystal} is derived under the
assumption that the wave is isotropically scattered by the
scattering center. This assumption is true, e.g.,  for  thermal
and cold neutrons, whose wavelength $\lambda$ is much larger than
the nuclear radius.
However, various artificial crystals (metamaterials), which  are
currently being studied, are  composed of the elements comparable
in size with the wavelength of the incident radiation (photons,
cold and ultra-cold neutrons).
As a consequence, the amplitude of wave scattering by the center
becomes dependent on the scattering angle. Moreover, the
scattering amplitude can depend on the scattering angle even when
the wavelength is large compared to the size of the scatterer;
this is true, e.g., for scattering of electromagnetic waves.
In this regard, one may wonder what the expression for the
refractive index will look like in this case, and what equation
will describe diffraction in crystals composed of such scatterers.

The refraction and diffraction of photons in artificial
(electromagnetic, photonic) crystals built from metallic wires
have been considered previously in
\cite{Baryshevsky,Gurnevich2010,Gurnevich2012, nuovo}.
The characteristic feature of electromagnetic wave scattering by a
wire is that the scattering amplitude is independent of the
scattering angle if the wave has a polarization parallel to the
wire and a wavelength $\lambda$ much larger than the wire's
radius, while  for wave polarization orthogonal to the wire, the
scattering amplitude $f$ depends on the scattering angle even for
the wavelengths $\lambda$ much larger than the wire's radius
\cite{Gurnevich2010,Gurnevich2012}
\begin{equation}
\label{EU}
 f=f_0+f_1\cos \theta.
\end{equation}
 A detailed analysis given in \cite{Gurnevich2010,Gurnevich2012} has shown that
in the case of angular dependence \eqref{EU}, the following
expression for the refractive index is valid:
\begin{equation}
n^2=1+\frac{4\pi}{k^2\Omega_3}\left\{\frac{f_0}{1+ikf_0}+\frac{f_1}{1+i\frac{k}{3}f_1}\right\}.
\label{eq:n2_3D_crystal_anisotropic1}
\end{equation}
According to \eqref{eq:n2_3D_crystal_anisotropic1}, in
the case of purely elastic scattering, multiple rescattering of
waves in crystals leads to the absence of an imaginary part in
$n^2$.

The present paper considers the refraction and diffraction of
waves  in three-dimensional 
crystals built from the centers scattering the incident wave
anisotropically. Here the results of \cite{Gurnevich2012} are
generalized to the case of arbitrary angular dependence
$f(\theta)$.
We derive the expression for the refractive index and the
equations describing dynamical diffraction of waves that
generalize the equations of dynamical diffraction of X-rays and
neutrons in crystals to this case.

\section{Scattering by a single center}
\subsection{Method of partial waves}

For concreteness, let us consider scattering of a scalar wave by a
spherically symmetric potential.
{Because the angular momentum in a spherically symmetric field is
conserved, it appears reasonable to present the initial plane wave
$\Psi_0(\vec{r})=e^{i\vec{k}\vec{r}}$ as a superposition of
spherical waves with different angular momenta (partial waves) and
consider scattering of each of these waves separately. Let us
recall here that the wave function of a state with angular
momentum  $l$ and the projection thereof
 $m$ can be expressed as \cite{landau77quant}
\[
\psi_{klm}=R_{kl}(r)Y_{lm}(\theta,\varphi),
\]
where $Y_{lm}$ are the spherical functions, and the radial
function $R_{kl}$ has the form
\begin{equation}
  R_{kl}(r)=2kj_l(kr),
  \label{eq:Rkl}
\end{equation}
where $j_l$ are the so-called spherical Bessel functions. Knowing
the asymptotic behavior of the Bessel functions, we find
$R_{kl}\approx \dfrac{2}{r}\sin (kr-\frac{\pi l}{2})$  far from
and $R_{kl}\approx \dfrac{2 k^{l+1}}{(2l+1)!!}r^l$ near the origin
of coordinates.

In place of the functions $R_{kl}$, the scattering theory often
uses the functions $R_{kl}^{\pm}(r)=\pm i k h_l^{(1,2)}(kr)$
corresponding to spherical waves converging towards (the minus
sign and the spherical Hankel function of the second kind) and
diverging from the center (the plus sign and the Hankel function
of the first kind). The limiting expressions for the functions
$R_{kl}^\pm$  at $kr\gg 1$ and near the origin of coordinates have
the form
\begin{eqnarray}
  R_{kl}^{\pm}\approx \frac{1}{r}e^{\pm i (kr-\pi l/2)}, & kr\gg 1;
  \label{eq:Rkl_limits_top}
  \\
  R_{kl}^{\pm}\approx  \frac{(2l-1)!!}{k^l}r^{-l-1}, & r\rightarrow 0,
  \label{eq:Rkl_limits_bottom}
\end{eqnarray}
where $(2l-1)!!=1\cdot 3 \cdot 5 \cdot ... \cdot (2l-1)$. }

Let an initial plane wave propagate along the $z$-axis in the
Cartesian coordinate system: $\Psi_0(\vec{r})= e^{i\vec{k}\vec{r}}
\equiv e^{ikz}$. The scatterer is located at the origin of
coordinates. Because the function $e^{ikz}$ is axially symmetric
about the $z$-axis, its partial-wave expansion contains only
spherical functions that are independent of the angle $\varphi$
(i.e., $m=0$) and has the form (as shown, for example, in
\cite{landau77quant}):
\begin{equation}
  e^{ikz} = \sum\limits_{l=0}^\infty (2l+1)i^l j_l(kr)P_l(\cos\theta)
  = -\frac{i}{2k}\sum\limits_{l=0}^{\infty}(2l+1)i^l(R_{kl}^+ - R_{kl}^-) P_l(\cos\theta),
  \label{eq:plane_wave_decompos}
\end{equation}
where $P_l$ are the Legendre polynomials.
In view of
 \eqref{eq:Rkl_limits_top}, we can easily see that  far from the
center, each partial wave in  \eqref{eq:plane_wave_decompos} is a
sum of two spherical waves:  diverging from  and converging
towards the center.

The solution of the scattering problem for $\Psi(\vec{r})$ can be
sought in a  similar form as in \eqref{eq:plane_wave_decompos}:
\begin{equation}
  \Psi(\vec{r}) = -\frac{i}{2k}\sum\limits_{l=0}^\infty (2l+1)i^l P_l(\cos\theta)R^*_l(r),
  \label{eq:solution_asympt}
\end{equation}
where  $R^*_l(r)$ is the  function of interest.
Since the
interaction between the incident wave (incident particle flow) and
the scattering potential will affect only the amplitude of
spherical waves diverging from the center, the asymptotic
expression for $R^*_l(r)$ at $kr\gg 1$ can be written in the form
\cite{davydov76quant}
\begin{equation}
  -\frac{i}{2k} R^*_l(r) = -\frac{i}{2k}\left( S_l R_{kl}^+ - R_{kl}^-\right) = j_l(kr) -
  \frac{i}{2k} (S_l-1) R_{kl}^+
  \approx \frac{1}{kr}\sin(kr-l\pi/2) +
  \frac{i}{2}(1-S_l)\frac{e^{i(kr-l\pi/2)}}{kr}.
  \label{eq:Rl}
\end{equation}
The coefficient $S_l$ is the diagonal matrix element of the
scattering matrix $S$\ and corresponds to the orbital angular
momentum $l$ \cite{landau77quant,davydov76quant}.

Thus the wave function $\Psi(\vec{r})$ at $kr\gg 1$ can finally be
presented as a sum of the incident plane wave and the scattered
spherical wave diverging from the center
\begin{equation}
  \Psi(\vec{r}) = \Psi_0(\vec{r}) + \Psi_{sc}(\vec{r}) = \Psi_0 + f(\theta)\frac{e^{ikr}}{r}.
  \label{eq:wavefunction_asympt}
\end{equation}
Using  \eqref{eq:plane_wave_decompos} and
\eqref{eq:Rl}-\eqref{eq:wavefunction_asympt}, one can easily find
that the scattering amplitude $f(\theta)$ is expressed in terms of
the  matrix elements of the scattering matrix as\footnote{The
convergence conditions for this series, see in
\cite{landau77quant}.}
\begin{equation}
  f(\theta) = \frac{i}{2k}\sum\limits_{l=0}^\infty (2l+1)(1-S_l)P_l(\cos\theta) \equiv
  \sum\limits_{l=0}^\infty f_l P_l(\cos\theta),
  \label{eq:amplitude_of_scattering}
\end{equation}
where the notation $f_l = \dfrac{i}{2k}(2l+1)(1-S_l)$ is
introduced. The scattered wave at $kr \gg 1$ has the form
\begin{equation}
  \Psi_{sc}(\vec{r}) = -\frac{i}{2k}\sum\limits_{l=0}^\infty (2l+1) i^l (S_l-1)R_{kl}^+(r)
  P_l(\cos\theta) = \sum\limits_{l=0}^\infty i^l f_l R_{kl}^{+} (r) P_l(\cos\theta) \approx
  f(\theta)\frac{e^{ikr}}{r} . 
  \label{eq:scattered_wave}
\end{equation}
 For further consideration it is convenient  to establish how the
amplitude of a partial wave, which acts directly on the scatterer,
relates to the amplitude of the resulting scattered wave.
This can easily be done by term-by-term comparison of the
expressions \eqref{eq:scattered_wave} for the scattered wave and
\eqref{eq:plane_wave_decompos} for the incident wave with the
limiting values  of $j_l(kr)$ as $r\rightarrow 0$ (i.e., near the
scatterer) substituted for the functions $j_l(kr)$.
As a result, we have that if  the partial wave
$\Psi_l^0=i^lk^lr^lP_l(\cos\theta)$ acts on the scatterer, then
 the scattered wave far from it has the form
 \begin{equation}
 \Psi_l^{sc}= (2l-1)!!\, i^l f_l R_{kl}^+(r) P_l(\cos\theta) \approx (2l-1)!! f_l
\frac{e^{ikr}}{r}P_l(\cos\theta).
\label{eq:partialwave_scattering}
\end{equation}

\subsection{Scattering matrix}

In the general case,  it is convenient for describing the
scattering process to introduce the operator $\hat{T}$ (the
so-called $T$-operator or $T$-matrix), whose matrix elements on
the energy surface are related to  the matrix elements of the
scattering matrix  $S$ as  \cite{davydov76quant}
\begin{equation}
  S_{ba} = \delta_{ba} - 2\pi i T_{ba} \delta (E_b - E_a),
  \label{eq:Tmatrix}
\end{equation}
where $E_{b}$ and  $E_{a}$ are the total energy of the final
 $b$ and the initial $a$ states, respectively. Let us recall here
 that the matrix elements of the $T$ matrix  are proportional to
 the scattering amplitude (or the reaction amplitude, if a
 reaction takes place). Particularly, partial amplitudes $f_l$ can be expressed
 in terms of matrix elements $T_l$ in a rather simple way
\begin{equation}
 f_l=-\frac{\pi}{k}(2l+1)T_l.
  \label{eq:AlTl}
\end{equation}

Sometimes it is more convenient to define the asymptotic behavior
of the wave function at infinity in the form of standing waves
\cite{Newton1969}, rather than in the form of a superposition of
converging and diverging waves as is done in \eqref{eq:Rl}
\begin{equation}
 R_l \sim \sin (kr-\frac{l\pi}{2}) + K_l \cos (kr-\frac{l\pi}{2}).
 \label{eq:Rl_Kasymtotic}
\end{equation}

In the standing-wave representation,  a new operator $\hat{K}$
acts as a scattering operator (the corresponding matrix is usually
called the reaction matrix). Like the operator $\hat{T}$,  this
operator allows one to solve  any scattering problem
\cite{Newton1969, davydov58nucl}.

Let us note that in contrast to $T$ and $S$, the reaction matrix
$K$, introduced here, is Hermitian ($K^\dag=K$), and so its
eigenvalues are real. Matrix elements of the operator $\hat{K}$ on
the energy surface are related to the matrix elements of the
$T$-matrix by the so-called Heitler  equations \cite{Goldberger,
davydov58nucl}
\begin{equation}
  K_{ba} = T_{ba} + i\pi \sum\limits_c K_{bc}T_{ca}\delta (E_c-E_a).
  \label{eq:KmatrixT}
\end{equation}
Relation  \eqref{eq:KmatrixT} takes rather a simple form if the
matrices $K$ and $T$ are diagonal -- for example, when the orbital
angular momentum is conserved, then  for every partial wave having
the moment $l$ we have
\begin{equation}
  K_l = \frac{T_l}{1 - i\pi T_l}.
  \label{eq:KlTlcoupling}
\end{equation}

\section{Propagation of waves in crystals}

Let us proceed to the consideration of wave scattering in
crystals. To begin with, we shall consider the case when the
diffraction conditions are not fulfilled, and a single refracted
wave propagates in the crystal.
%
%
Following \cite{Gurnevich2012}, we assume that the amplitude of
scattering by a single center is small, $k|f(\theta)|\ll 1$, and
the refractive index is close to unity,\footnote{In the general
case, the denominator of the expression for the effective
scattering amplitude in \eqref{eq:n2_3Dcrystal} should have the
form $1+iC_1(k)A + C_2(k)A$, where $C_1(k)$ and $C_2(k)$ are the
real coefficients (see e.g. \cite{lax2}). If $|n^2-1|\ll 1$, then
multiple rescattering of waves by the centers composing the
crystal has a significant influence only on the imaginary part of
the refractive index, having practically no effect on its real
part, and we only need to find  $C_1(k)$. If $n$ differs
significantly from unity, we need to calculate both $C_1(k)$ and
$C_2(k)$. The relevant example can also be found in
\cite{Baryshevsky, Belov1} that discuss a two-dimensional crystal
built from metallic threads (wires). In a long-wave approximation,
the refractive index of such a crystal can be appreciably
different from unity. In the present paper we shall focus only on
calculating $C_1(k)$. } $|n^2-1|\ll 1$.

\begin{figure}[h]
    \begin{minipage}[h]{0.49\linewidth}
        \center{\includegraphics{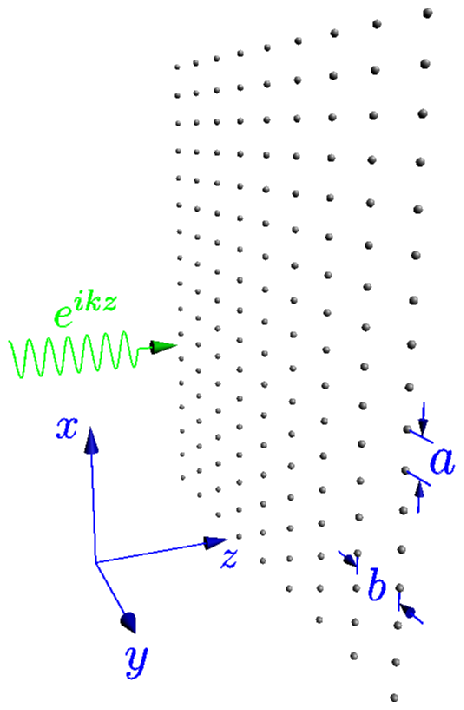} \\ a)}
    \end{minipage}
    \begin{minipage}[h]{0.49\linewidth}
        \center{\includegraphics{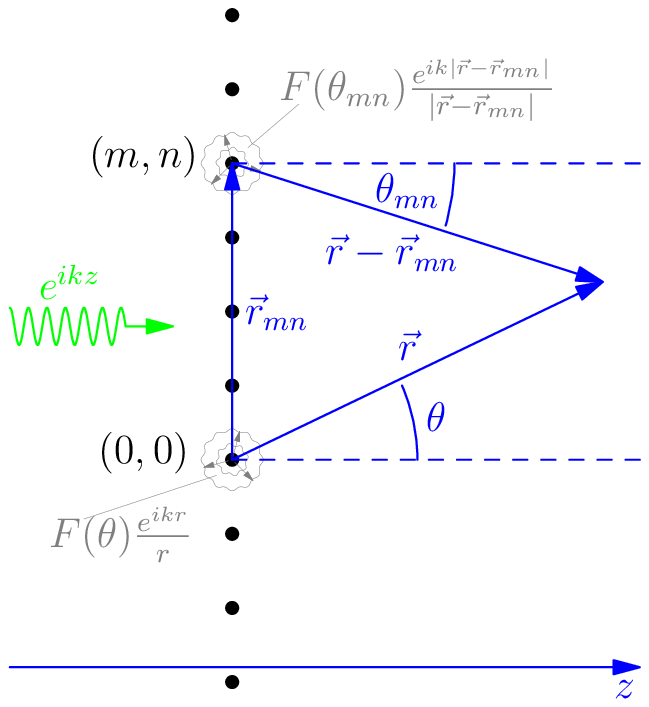} \\ b)}
    \end{minipage}
    \caption{Plane-wave scattering by a two-dimensional  grating: a) 3D view; b) relation between the polar angles  $\theta$ and $\theta_{m,n}$.}
    \label{fig:spheres}
\end{figure}

We shall first analyze scattering by one crystal plane. Let a
plane wave $\Psi_0 = e^{ikz}$ be normally incident on a
two-dimensional grating composed of scattering centers (see Fig.~\ref{fig:spheres}).
Let us specify the coordinates of the scatterers in the form
$\vec{r}_{m,n}=(x_{m,n},y_{m,n},z_{m,n}) = (ma,nb,0)$, where
 $m,n$ are the integers, while  $a$ and $b$ are the grating periods.
 Multiple scattering is (for the moment) ignored. Then the wave $\Psi(\vec{r})$
 that has passed through the grating can be written  as a sum of
 the incident plane wave and spherical waves scattered with the
 amplitude $f(\theta)$
\begin{equation}
  \Psi(\vec{r}) = e^{ikz} + ik \sum\limits_{m,n,l}i^l f_l h_l(k|\vec{r}-\vec{r}_{m,n}|)
  P_l(\cos\theta_{m,n}),
  \label{eq:transmittedwave1}
\end{equation}
where $\theta_{m,n}$ is the angle between vectors $\vec{k}$ and
$\vec{r}-\vec{r}_{m,n}$. This sum can be transformed using the
Poisson summation formula. Particularly, at $l=0$ the summation
over  $m$ and $n$ yields
\begin{equation}
  ik \sum\limits_{m,n} h_0(k|\vec{r}-\vec{r}_{m,n}|) = \sum\limits_{m,n}\frac{e^{ik|\vec{r}-\vec{r}_{m,n}|}}{|\vec{r}-\vec{r}_{m,n}|} =
  \frac{2i\pi}{kab}e^{ik|z|} + \frac{1}{ab}\sum\limits_{m,n\neq(0,0)}\Phi\left(
  \frac{2\pi m}{a},\frac{2\pi n}{b}\right),
  \label{eq:sum1}
\end{equation}
where the function $\Phi(\alpha,\beta)$ is defined as follows
\begin{equation}
  \Phi(\alpha,\beta)=\begin{cases}
    \dfrac{2i\pi e^{-i(\alpha x + \beta y)}}{\sqrt{k^2-\alpha^2-\beta^2}}
    e^{i\sqrt{k^2-\alpha^2-\beta^2}|z|},&\text{when $k^2>\alpha^2+\beta^2$;}\\
    \dfrac{2\pi e^{-i(\alpha x + \beta y)}}{\sqrt{\alpha^2+\beta^2-k^2}}
    e^{-\sqrt{\alpha^2+\beta^2-k^2}|z|},&\text{when $k^2<\alpha^2+\beta^2$.}
  \end{cases}
  \label{eq:Phi_function}
\end{equation}
For simplicity, let us assume that the conditions $ka<2\pi$ and
$kb<2\pi$ are fulfilled. Then the second term in \eqref{eq:sum1}
appears to be a strongly attenuating wave, and at large values of
$z$ (as a rule, when $z$ becomes $z\sim a,b$), we obtain that a
plane wave with the amplitude $A=\left(1 + \frac{2i\pi
f_0}{kab}\right)$ is produced through the interaction between a
plane wave with unit amplitude $e^{ikz}$ and a plane grating.

Let us note that at large values of $k$, instead of one plane wave
produced through scattering,  several plane waves with wave
vectors pointing in different directions may appear. When, for
example, $2\pi<ka<4\pi$ and  $kb<2\pi$,  it follows from
\eqref{eq:sum1}-\eqref{eq:Phi_function} that the resulting field,
induced through scattering by a plane grating, is a superposition
of three plane waves with wave vectors $k\vec{e}_z$,
$\frac{2\pi}{a}\vec{e}_x + \sqrt{k^2-(2\pi/a)^2}\vec{e}_z$, and
$-\frac{2\pi}{a}\vec{e}_x + \sqrt{k^2-(2\pi/a)^2}\vec{e}_z$
($\vec{e}_x$ and $\vec{e}_z$ are the versors of a Cartesian
coordinate system). However, by means of simple but cumbersome
calculations (step-by-step consideration of the cases
$2\pi<ka<4\pi$, $4\pi<ka<6\pi$, $2\pi<kb<4\pi$, etc.) we can
rigorously show that taking account of additional
 waves has no influence on the final result for
the crystal refractive index
 and it remains valid for any values of $k$ beyond the conditions of
 diffraction.

Knowing the value of the sum over $m$ and $n$ at $l=0$ in
\eqref{eq:transmittedwave1}, we can calculate the sums for
$l\neq0$ using the well-known recurrent relations  for Legendre
polynomials and spherical Hankel functions \cite{Janke, Mors1953}.
We can readily show that taking sums over $m$ and $n$ for all $l$
gives the already known result $\frac{2i\pi}{kab}e^{ikz}$; then we
finally obtain the following expression for the wave that has
passed through the grating:
\begin{equation}
  \Psi(\vec{r}) \approx e^{ikz} + \frac{2i\pi}{kab}e^{ikz} \sum\limits_l f_l =
  \left(1 + \frac{2i\pi}{kab}f(0)\right)e^{ikz},
  \label{eq:transmittedwave2}
\end{equation}
where $f(0)$ is the zero-angle scattering amplitude. Analyzing the
successive scattering of waves by several ($m$ number) gratings
spaced a distance $c$ apart, we obtain the following expression
for the transmitted wave:
\[\Psi(\vec{r})=\left(1 +
\frac{2i\pi}{kab}f(0)\right)^me^{ikz}.
\]
It can be presented in the form $\Psi(\vec{r})=e^{iknz}$, where
the refractive index $n$ is defined by the equality (valid when
$|n-1|\ll 1$)
\begin{equation}
  n \approx 1 + \frac{2\pi}{k^2\Omega_3}f(0),
  \label{eq:nwithoutrescattering}
\end{equation}
where $\Omega_3=abc$ is the unit cell volume of the crystal. This
equality coincides with the expression for $n$ in the case of
chaotically distributed scatterers \cite{BVG95Nuclearoptics}.

In the above analysis  we ignored the influence of multiple
rescattering of waves on the scattering process. In actuality,
each center scatters not only the initial plane wave, but also
spherical waves, which have been scattered by all other centers.
Let us take into consideration this multiple rescattering of waves
by the centers composing a crystal. Let us first analyze the
process of rescattering by one crystal plane. Obviously, in the
general case the wave is the sum of the initial incident wave and
spherical waves, which diverge from each center. But, unlike the
case described in \eqref{eq:transmittedwave1}, their amplitudes
will differ from the amplitude $f(\theta)$ of scattering by a
single center.

Let $F(\theta)$  denote the amplitude of scattering by the center
that is a part of a crystal plane. In a similar manner as we
considered the amplitude $f(\theta)$, we can expand the amplitude
$F(\theta)$ using Legendre polynomials and write $F(\theta)$ in
the form $F(\theta)=\sum\limits_l F_lP_l(\cos\theta)$. Since the
grating is infinite and periodic in the ($x$,$y$) plane, the
scattering amplitude $F(\theta)$ is independent of the position of
the scatterer $(m,n)$ in this plane. As a consequence, we only
need to find the value of $F(\theta)$  for the scatterer located
at the origin of coordinates (for brevity's sake it will be termed
``reference scatterer'').

To obtain the equation for $F(\theta)$, let us note that there are
two waves scattered by the reference center: the initial plane wave
$\Psi_0=e^{ikz}$ and the wave
\begin{equation*}
 \Psi_{sc, 0}=
ik\sum\limits_{(m,n)\neq(0,0),l} i^l F_l
h_l(k|\vec{r}-\vec{r}_{m,n}|) P_l(\cos\theta_{m,n}),
\end{equation*}
coming from all other centers. As a result of scattering of these
two waves, a diverging spherical wave of amplitude $F(\theta)$ is
produced.

To analyze the scattering process using the method described in
section 2, we only need to properly perform the partial-wave
expansion of the wave $\Psi_{sc, 0}$. Let us consider a wave
scattered by the center $(m,n)$:
\begin{equation*}
  \Psi_{sc}^{(m,n)}=ik \sum\limits_{l=0}^{\infty} i^l F_l h_l(k|\vec{r}-\vec{r}_{m,n}|)
  P_l(\cos\theta_{m,n}).
\end{equation*}
We can see that for the angles $\theta_{m,n}$ and  $\theta$, the
following relation holds  (see Fig.~\ref{fig:spheres}, b)):
\begin{equation}
  \cos\theta_{m,n} = \frac{r}{|\vec{r}-\vec{r}_{m,n}|}\cos\theta.
  \label{eq:cos_theta}
\end{equation}
Then in the vicinity of the reference scatterer  ($r\rightarrow
0$), the wave  $\Psi_{sc}^{(m,n)}$ has the form
\begin{equation}
  \Psi_{sc}^{(m,n)}=ik \sum\limits_{j=0}^{\infty} i^j F_j h_j(kr_{m,n})
  P_j\left(\frac{r}{r_{m,n}}\cos\theta\right).
  \label{eq:psi_incident_mn}
\end{equation}
Legendre polynomials of argument  $\cos\theta\cdot r/r_{m,n}$ can
be expanded in Legendre polynomials of argument $\cos\theta$
\begin{equation}
  P_j\left(\frac{r}{r_{m,n}}\cos\theta\right) = \sum\limits_{l=0}^{\infty}
  \frac{1}{2}\alpha_{lj} P_l(\cos\theta),
  \label{eq:Legendre_decompos}
\end{equation}
where
\begin{equation}
  \alpha_{lj} = (2l+1)\int\limits_{0}^{\pi}P_j\left(\frac{r}{r_{m,n}}\cos\theta\right)
  P_l(\cos\theta)\sin\theta d\theta.
  \label{eq:alpha_lj}
\end{equation}
Upon evaluating the integrals \eqref{eq:alpha_lj} and substituting
the obtained values  into \eqref{eq:psi_incident_mn} and
\eqref{eq:Legendre_decompos}, we get the following expression for
the wave $\Psi_{sc}^{(m,n)}$, which holds true when $r\rightarrow
0$ (for details, see Appendix~A):
\begin{equation}
  \Psi_{sc}^{(m,n)} \approx ik \sum\limits_{l=0}^\infty
  i^l r^l P_l(\cos\theta)\frac{1}{2^l}\frac{1}{(2l-1)!!}\left\{
    \frac{1}{r_{m,n}^l}\sum\limits_{j=0}^\infty F_{l+2j}h_{l+2j}(kr_{m,n})\frac{1}{2^{2j}}
    \frac{(2(l+j))!}{j!(l+j)!}
  \right\}.
  \label{eq:psi_incident_mn_decompos1}
\end{equation}
Now we only need to perform the summation of
\eqref{eq:psi_incident_mn_decompos1} over all  $m$ and  $n$ but
$(m,n)=(0,0)$.  Since we are interested in the correction to the
imaginary part of the scattering amplitude, it suffices to
 find only the real parts of the  sums $\sum\limits_{m,n}
h_{l+2j}(kr_{m,n})r_{m,n}^{-l}$ (as one can see in what follows).
After quite a cumbersome summation procedure (see Appendix A), we
finally have
\begin{equation}
  \Psi_{sc, 0} = \sum\limits_{(m,n)\neq(0,0)} \Psi_{sc}^{(m,n)} \approx
  ik \sum\limits_{l=0}^\infty i^l r^l k^l P_l(\cos\theta)\frac{1}{(2l-1)!!}
  \left\{ -\frac{1}{2l+1}F_l + \frac{2\pi}{k^2 a b}\sum\limits_{j=0}^\infty F_{l+2j} \right\}.
  \label{eq:psi_incident_mn_decompos2}
\end{equation}
In view of  \eqref{eq:plane_wave_decompos} and
\eqref{eq:psi_incident_mn_decompos2}, the wave acting on the
reference scatterer has the form
\begin{equation}
  \Psi_{inc} = e^{ikz} + \Psi_{sc, 0} =
  \sum\limits_{l=0}^\infty \frac{1}{(2l-1)!!}i^l r^l k^l P_l(\cos\theta)
  \left\{ 1 - ik\frac{F_l}{2l+1} + \frac{2\pi i}{kab}\sum\limits_{j=0}^\infty F_{l+2j}
  \right\}.
  \label{eq:incident_wave_resc}
\end{equation}
The interaction between this wave and the scatterer results in the
formation of a scattered wave, which, according to
\eqref{eq:partialwave_scattering}, can be written as follows:
\begin{equation}
  \Psi_{sc}^{(0,0)} =
  \sum\limits_{l=0}^\infty i^l f_l R_{kl}^+(r) P_l(\cos\theta)
  \left\{ 1 - ik\frac{F_l}{2l+1} + \frac{2\pi i}{kab}\sum\limits_{j=0}^\infty F_{l+2j}
  \right\}.
  \label{eq:scattered_wave_0}
\end{equation}
On the other hand, the wave scattered by the reference center is
just a diverging spherical wave of amplitude   $F(\theta)$
\begin{equation}
  \Psi_{sc}^{(0,0)} = \sum\limits_{l=0}^\infty i^l F_l R_{kl}^+(r) P_l(\cos\theta).
  \label{eq:scattered_wave_1}
\end{equation}
Setting \eqref{eq:scattered_wave_0} equal to
\eqref{eq:scattered_wave_1}, we can obtain the following set of
equations for partial amplitudes  $F_l$:
\begin{equation}
   F_l = f_l - \frac{ik}{2l+1}f_lF_l + \frac{2i\pi}{kab}f_lF_l +
   \frac{2i\pi}{kab}f_l\sum\limits_{j=1}^\infty F_{l + 2j}.
  \label{eq:equations_Fl}
\end{equation}

Assuming that the scattering amplitude is small  ($kf_l\ll 1$), we
can solve the equations and obtain the following expressions for
the amplitudes $F_l$:
\begin{equation}
  F_l \approx \frac{f_l}{1 + \frac{ik}{2l+1}f_l - \frac{2i\pi}{kab}f_l}.
  \label{eq:Fl_values}
\end{equation}

Thus when the wave $\Psi_0(\vec{r})=e^{ikz}$ is scattered by a
plane grating of scattering centers, a plane wave is produced (at
$z>0$), whose amplitude can be described by the expression that
follows  rather than by \eqref{eq:transmittedwave2}
\begin{equation}
 \Psi(\vec{r})=e^{ikz} +
\frac{2i\pi}{kab}e^{ikz}\sum\limits_{l=0}^\infty F_l.
  \label{eq:transmitted_plus}
\end{equation}
When $z<0$, the wave field induced through scattering is a sum of
the incoming wave incident on the grating and the wave reflected
from it
\begin{equation}
 \Psi(\vec{r})=e^{ikz} +
\frac{2i\pi}{kab}e^{-ikz}\sum\limits_{l=0}^\infty (-1)^l F_l.
  \label{eq:transmitted_minus}
\end{equation}
The further analysis is performed in a similar manner as in
\cite{Gurnevich2012}. It is convenient to introduce the following
notation: $F^*_0\equiv \sum\limits_{l=0}^\infty F_{2l}$,
$F^*_1\equiv \sum\limits_{l=0}^\infty F_{2l+1}$. Then
\eqref{eq:transmitted_plus}-\eqref{eq:transmitted_minus} can be
written in a more compact form
\begin{equation}
 \Psi(\vec{r})=e^{ikz} +
\frac{2i\pi}{kab}(F^*_0 \pm F^*_1)e^{ik|z|},
\label{eq:transmitted_both}
\end{equation}
where the plus sign refers to the case $z>0$, while the minus sign
refers to $z<0$ (forward and backward scattering, respectively)

Worthy of mention is that \eqref{eq:transmitted_both} has the same
form as in the analysis of the simplest case of anisotropic
scattering ($l=0,1$) with the amplitude  $F^*(\theta)=F^*_0 +
F^*_1\cos\theta$. The quantities {${\dfrac{2i\pi}{kab}(F^*_0 \pm
F^*_1)}$} can be interpreted as the ``amplitudes of forward
(backward) scattering''  of a plane wave by a two-dimensional
grating (crystal plane) composed of scattering centers.

Let us consider a crystal composed of a set of above-discussed
plane two-dimensional gratings placed at a distance $c$ apart. The
wave incident on the grating placed at the origin of coordinates
is a sum of waves scattered by all other gratings. Following
\cite{Gurnevich2012}, let us assume that these waves have the
amplitude $\Phi = (\Phi_0 \pm \Phi_1)e^{iqcm}$, where $q$ is the
wave number in the crystal, $m$ is the integer (the number of the
grating), and neither $\Phi_0$ nor  $\Phi_1$  depends on $m$,
owing to the periodicity of the crystal lattice. Because all these
waves are scattered by this grating with the known amplitude
$\dfrac{2i\pi}{kab}(F^*_0 \pm F^*_1)$ and produce a plane wave of
amplitude $(\Phi_0 \pm \Phi_1)$, we can write the following set of
equations for the amplitudes $\Phi_0$ and $\Phi_1$ (for details
see \cite{Gurnevich2012}):
\begin{equation}
  \Phi_0  =  F^*_0 \left\{ \frac{2\pi i}{kab} \Phi_0 S_3 + \frac{2\pi i}{kab} \Phi_1 S_4 \right\}, \\
  \label{eq:phi0}
  \end{equation}
  \begin{equation}
  \Phi_1  =  F^*_1 \left\{ \frac{2\pi i}{kab} \Phi_0 S_4 + \frac{2\pi i}{kab} \Phi_1 S_3 \right\},
  \label{eq:phi1}
\end{equation}
where the sums $S_3$ and $S_4$ are
\begin{eqnarray*}
  S_3 = \sum\limits_{m\neq 0} e^{iqcm}e^{ikc|m|} = -1 -i \frac{\sin kc}{\cos kc - \cos qc} \simeq -1 -
  \frac{2i}{kc}\cdot \frac{1}{n^2 -1}, \\
  S_4 = \sum\limits_{m = 1}^{\infty} (e^{-iqcm} - e^{iqcm})e^{ikcm} = -i \frac{\sin qc}{\cos kc - \cos qc}
  \simeq  -\frac{2i}{kc}\cdot \frac{1}{n^2 -1}.
\end{eqnarray*}
By setting the determinant of \eqref{eq:phi0}-\eqref{eq:phi1}
equal to zero, we obtain the dispersion equation for the wave
number $q$. The solution of this dispersion equation has the form
\begin{equation}
  n^2 = \frac{q^2}{k^2} = 1 + \frac{4\pi}{k^2\Omega_3}\sum\limits_l \frac{f_l}{1 + \frac{ik}{2l+1}f_l},
  \label{eq:refractive_index1}
\end{equation}
where  $\Omega_3$ is the unit cell volume of the crystal.

As is seen, in the case of isotropic scattering (only $l=0$ is
considered), this equation reduces to a well-known formula for the
crystal refractive index:
$n^2=1+\dfrac{4\pi}{k^2\Omega_3}\dfrac{f_0}{1+ikf_0}$
\cite{BVG95Nuclearoptics,Baryshevsky1966}, while in the simplest
anisotropic case ($l=0,1$) we arrive at equation
\eqref{eq:n2_3D_crystal_anisotropic1}, first derived in
\cite{Gurnevich2012}.

Remembering  the relationship  \eqref{eq:AlTl}  between  the
partial amplitudes $f_l$ and the matrix elements $T_l$, we can
present \eqref{eq:refractive_index1} in the form
\begin{equation}
  n^2 = 1 - \frac{4\pi^2}{k^3\Omega_3}\sum\limits_l (2l+1)\frac{T_l}{1 - i\pi T_l},
  \label{eq:refractive_index_T}
\end{equation}
It follows from \eqref{eq:KlTlcoupling} that the expression for
$n^2$ can be recast as
\begin{equation}
   n^2 = 1 - \frac{4\pi^2}{k^3\Omega_3}\sum\limits_l (2l+1)K_l.
  \label{eq:refractive_index_K}
\end{equation}
Let us recall that if the  scatterers make up a chaotic medium,
then
\begin{equation}
  n^2 = 1 + \frac{4\pi\rho}{k^2}f(0) = 1 - \frac{4\pi^2\rho}{k^3}\sum\limits_l (2l+1)T_l,
  \label{eq:refractive_index_ch}
\end{equation}
where $\rho$ is the density of scatterers in the medium (in
crystals $\rho=\frac{1}{\Omega_3}$).

As we see, what differs  \eqref{eq:refractive_index_K}, defining
the crystal refractive index,  from \eqref{eq:refractive_index_ch}
is the reaction matrix $K_l=\dfrac{T_l}{1-i\pi T_l}$ that appears
in \eqref{eq:refractive_index_K} instead of the scattering matrix
$T_l$ in \eqref{eq:refractive_index_ch}. This means that at a
large distance from the  center located inside the crystal,
multiple coherent rescattering of waves in crystals changes  the
asymptotic behavior of a  wave function  according to
\eqref{eq:Rl_Kasymtotic}. But at a large distance from the
crystal, the asymptotic behavior of a wave function still has the
form
\begin{equation}
\mathcal{F}\frac{e^{ikr}}{r},
\end{equation}
where $\mathcal{F}$ is the scattering amplitude as a whole.

 The basic difference
 between \eqref{eq:refractive_index_K} and  \eqref{eq:refractive_index_ch}
  is that in the case of elastic scattering by a
 single center, equation \eqref{eq:refractive_index_K} is purely real,
 while \eqref{eq:refractive_index_ch} has both real and imaginary
 parts (the wave attenuates in the medium).  Equation
\eqref{eq:refractive_index_K} is real because in the case of
elastic scattering, the matrix $K$  is Hermitian  and  diagonal in
$l$. Thus, wave attenuation in crystals may come only from
inelastic scattering by the centers composing the crystal, while
in a chaotic medium, both  inelastic scattering by individual
centers and purely elastic scattering contribute to the
attenuation of waves.

Let us take as an example the electromagnetic crystal built from
parallel metallic wires (see
\cite{Baryshevsky,Gurnevich2010,Gurnevich2012}). As shown in
\cite{Gurnevich2010},   the Vavilov-Cherenkov effect in such
crystals can  be observed  because the refractive index of a wave
with orthogonal polarization is greater than unity.  Let the wire
spacing in the crystal be $d_x=d_y= 0.1$~cm and the wire radius
$R=25$~$\mu$~m. Using well-known expressions for the amplitude of
electromagnetic wave scattering by a wire \cite{Baryshevsky}, we
can show that, e.g., at a frequency $f=1$~THz,  the  refractive
index of a crystal is $n^2\approx 1.002$, while for a chaotic
medium of the same density $n^2\approx 1.002 + 1.3\cdot10^{-3}i$.
Let a  crystal (or a chaotic medium) have a thickness  $z=10$~cm.
We can readily obtain that in a chaotic medium of such thickness,
the amplitude of a transmitted electromagnetic wave of frequency
$1$~THz will reduce by a factor of $\exp(kz\mathrm{Im}n)\approx
3.9$ (i.e., the reduction is significant), while in a crystal it
will remain practically the same. It can be seen that the use of
\eqref{eq:refractive_index_ch} for calculating the refractive
index of crystals  could lead to the erroneous conclusion that the
generated Cherenkov radiation
 rapidly decays in crystals.

These results (the fact that scattering by the center belonging to
the crystal is described by $\dfrac{T_l}{1-i\pi T_l}$ rather than
by the matrix elements $T_l$) should necessarily be taken into
account in considering diffraction in crystals. Dynamical
diffraction of X-rays and neutrons in crystals can be described by
the following set of equations
\cite{BVG95Nuclearoptics,Batterman1964,Pinsker,Lamb}:
\begin{eqnarray}
 \left( 1 - \frac{k^2}{k_0^2} \right) \varphi(\vec{k})
  + \sum\limits_{\vec{\tau}} g(\vec{\tau})
  \varphi(\vec{k}-\vec{\tau}) =  0,
  \label{eq:dynamical_system1}
  \\
  \Psi(\vec{r})  =  \sum\limits_{\vec{\tau}} \varphi(\vec{k}
  + \vec{\tau}) e^{i(\vec{k}+\vec{\tau})\vec{r}},
  \label{eq:dynamical_system2}
  \end{eqnarray}
where
$g(\vec{\tau})=\dfrac{4\pi}{k^2\Omega_3}f(\vec{k},\vec{k}+\vec{\tau})$
  is the structure amplitude and $\vec{\tau}$ is the reciprocal lattice vector of the
  crystal.

The elements of the scattering matrix $T_l$ in
\eqref{eq:dynamical_system1}-\eqref{eq:dynamical_system2} appear
only in expression for the structure amplitude $g(\vec{\tau})$
(because they enter into the scattering amplitude
$f(\vec{k},\vec{k}+\vec{\tau})\equiv f(\theta_\tau)$). Replacing
the amplitude $T_l$ by $\dfrac{T_l}{1-i\pi T_l}$ gives the
following expression for the structure amplitude:
\begin{equation}
 g(\vec{\tau}) = \dfrac{4\pi}{k^2 \Omega_3}
 \sum\limits_l \frac{f_l}{1 + \frac{ik}{2l+1}f_l}P_l(\cos(\vec{k},\vec{k}+\vec{\tau}))=
-\dfrac{4\pi^2}{k^3 \Omega_3} \sum\limits_l (2l+1) \frac{T_l}{1
-i\pi T_l} P_l(\cos(\vec{k},\vec{k}+\vec{\tau})).
 \label{eq:structure_ampl}
 \end{equation}
The analysis shows that equations
\eqref{eq:dynamical_system1}--\eqref{eq:dynamical_system2}
together with \eqref{eq:structure_ampl} fully describe the
dynamical diffraction of waves in a crystal.

\section{Conclusion}

This paper considers the influence of multiple rescattering on the
propagation of waves in crystals composed of anisotropically
scattering centers. We used the  partial wave expansion method to
generalize the results of  \cite{Gurnevich2012}, valid only in a
particular case when the  amplitude of scattering by a single
center depends on the angle as $f(\theta)=f_0+f_1\cos \theta$
($s$- and $p$-scattering), to the case of arbitrary scattering
anisotropy (with orbital quantum numbers $l>1$). What is more, we
analyzed a general three-dimensional problem instead of a
two-dimensional one,  considered in \cite{Gurnevich2012}. It was
shown for the first time that the expression for the refractive
index of a three-dimensional crystal must include the quantities
$\dfrac{T_l}{1-i\pi T_l}$ (in the case of elastic scattering, they
coincide with the diagonal elements of  the reaction matrix
$\mathbf{K}$ [see \eqref{eq:KlTlcoupling}]) instead of the
diagonal elements of the scattering matrix $T_l$, which enter into
the equation in the case of a chaotic medium. As a consequence,
the structure amplitude in equations
\eqref{eq:dynamical_system1}--\eqref{eq:dynamical_system2}
describing the dynamical diffraction of waves in a crystal must be
calculated by formula \eqref{eq:structure_ampl}.

 Because a general
approach is applied to the description of the scattering process,
the results thus obtained are valid for a wide range of problems
and beneficial for many applications. These results can be of
interest for research into, e.g., diffraction of cold neutrons and
photons in crystals, nanocrystalline materials, as well as for the
description of  parametric and diffraction radiation in
electromagnetic crystals formed by anisotropically scattering
centers.


\begin{thebibliography}{10}

\bibitem{lax}
M.~Lax, ``Multiple scattering of waves,'' {\em Rev. Mod. Phys.},
vol.~23,
  pp.~287--310, Oct 1951.

\bibitem{Goldberger}
M.~Goldberger and K.~Watson, {\em Collision Theory}.
\newblock Structure of matter series, Wiley, New York, 1964.

\bibitem{BVG95Nuclearoptics}
V.~Baryshevsky, {\em High-Energy Nuclear Optics of Polarized
Particles}.
\newblock World Scientific, Singapore, 2012.

\bibitem{landau77quant}
L.~Landau and E.~Lifshitz, {\em Quantum Mechanics:
Non-Relativistic Theory}.
\newblock Statistical Physics, Pergamon Press, Oxford, 1977.

\bibitem{goldberger2}
M.~L. Goldberger and F.~Seitz, ``Theory of the refraction and the
diffraction
  of neutrons by crystals,'' {\em Phys. Rev.}, vol.~71, pp.~294--310, Mar 1947.

\bibitem{Batterman1964}
B.~Batterman and H.~Cole, ``Dynamical diffraction of x-rays by
perfect
  crystals,'' {\em Reviews of Modern Physics}, vol.~36, no.~3, pp.~681--717,
  1964.

\bibitem{lax2}
M.~Lax, ``Multiple scattering of waves ii. the effective field in
dense
  systems,'' {\em Phys. Rev.}, vol.~85, pp.~621--629, Feb 1952.

\bibitem{Baryshevsky1966}
V.~Baryshevskii, ``Neutron diffraction in a polarized crystal,''
{\em Sov.
  Phys. JETP}, vol.~24, no.~5, pp.~1068--1070, 1967.

\bibitem{Baryshevsky}
V.~Baryshevsky and A.~Gurinovich, ``Spontaneous and induced
parametric and
  smith-purcell radiation from electrons moving in a photonic crystal built
  from the metallic threads,'' {\em Nuclear Inst. and Meth. B}, vol.~252,
  no.~1, pp.~92 -- 101, 2006.

\bibitem{Gurnevich2010}
V.~Baryshevsky and E.~Gurnevich, ``The possibility of cherenkov
radiation
  generation in a photonic crystal formed by parallel metallic threads,'' {\em
  in Proc. 2010 International Kharkov Symposium on Physics and Engineering of
  Microwaves, Milimeter and Submilimeter Waves (MSMW'10), Kharkiv, Ukraine
  (2010)}, pp.~1--3.

\bibitem{Gurnevich2012}
V.~Baryshevsky and E.~Gurnevich, ``Dynamical diffraction theory of
waves in
  photonic crystals built from anisotropically scattering elements,'' {\em
  Journal of Nanophotonics}, vol.~6, no.~1, p.~061713, 2012.

\bibitem{nuovo}
V.~Baryshevsky, A.~Gurinovich, E.~Gurnevich, and A.~Lobko,
``Generation of
  medical x-ray and thz beams of radiation using table-top accelerators,'' {\em
  Il Nuovo Cimento C}, vol.~034, no.~4, pp.~199--205, 2011.

\bibitem{davydov76quant}
A.~Davydov, {\em Quantum Mechanics}.
\newblock International series in natural philosophy, Pergamon Press, New York,
  1976.

\bibitem{Newton1969}
R.~Newton, {\em Scattering Theory of Waves and Particles}.
\newblock Dover Publications, 1982.

\bibitem{davydov58nucl}
A.~Davydov, {\em Theory of the atomic nucleus [in Russian]}.
\newblock Fizmatgiz, Moscow, 1958.

\bibitem{Belov1}
P.~A. Belov, S.~A. Tretyakov, and A.~J. Viitanen, ``Dispersion and
reflection
  properties of artificial media formed by regular lattices of ideally
  conducting wires,'' {\em J. of Electromagn. Waves and Appl.}, vol.~16, no.~8,
  pp.~1153--1170, 2002.

\bibitem{Janke}
E.~Jahnke, F.~Emde, and F.~L{\"o}sch, {\em Tables of Higher
Functions}.
\newblock B.~G.~Teubner, Stuttgart, 1966.

\bibitem{Mors1953}
P.~M. Morse and H.~Feshbach, {\em Methods of Theoretical Physics}.
\newblock Mc Graw Hill, New York, 1953.

\bibitem{Pinsker}
Z.~Pinsker, {\em Dynamical Scattering of X-Rays in Crystals}.
\newblock Springer Series in Solid-State Sciences, Springer-Verlag, Berlin,
  1978.

\bibitem{Lamb}
W.~E. Lamb, ``Capture of neutrons by atoms in a crystal,'' {\em
Phys. Rev.},
  vol.~55, pp.~190--197, Jan 1939.

\bibitem{prudnikov1986integrals}
A.~Prudnikov, I.~Brychkov, and O.~Mari{\v{c}}ev, {\em Integrals
and Series:
  Special functions. Vol.2}.
\newblock CRC Press, 1986.

\bibitem{bateman1953v1}
H.~Bateman and A.~Erdelyi, {\em Higher transcendental functions,
vol.~1}.
\newblock Mc Graw-Hill, NY, 1953.

\end{thebibliography}


\newpage
\appendix
\section{Partial-Wave Expansion of  $\Psi_{sc, 0}$ }

Using \eqref{eq:Legendre_decompos}-\eqref{eq:alpha_lj}, let us
transform \eqref{eq:psi_incident_mn} for the wave
$\Psi_{sc}^{(m,n)}$ scattered by the center   $(m,n)$. First, we
find the coefficients $\alpha_{lj}$
\begin{equation}
  \alpha_{lj} = (2l+1)\int\limits_{-1}^{1}P_j(\frac{r}{r_{m,n}}x)P_l(x)dx.
  \label{eq:alpha_lj_2}
\end{equation}
Because  $P_j(rx/r_{m,n}$) is the $j$-th degree polynomial of
argument $rx/r_{m,n}$
\begin{equation*}
  P_j\left(\frac{r}{r_{m,n}}x\right)=\sum\limits_{k=0}^j a_k^{(j)}\left( \frac{r}{r_{m,n}}\right)^k x^k,
\end{equation*}
the quantities $\alpha_{lj}$ [according to
\eqref{eq:alpha_lj_2}] in the general case are  the $j$-th degree
polynomials of argument $r/r_{m,n}$
\begin{equation*}
  \alpha_{lj}=\sum\limits_{k=0}^j c_k \left( \frac{r}{r_{m,n}}\right) ^k.
\end{equation*}
Since we are concerned with the behavior of the wave
$\Psi_{sc}^{(m,n)}$ in the vicinity of the reference scatterer,
i.e., as   $r\rightarrow 0$, we shall keep only the first nonzero
term  in the power series expansion of $\alpha_{lj}$ over
$r/r_{m,n}$. We shall also take into account that  the integral
\begin{equation*}
  \int\limits_{-1}^1 P_l(x) x^m = 0
\end{equation*}
at all  $m<l$ (see~\cite{prudnikov1986integrals}).         
Then we have that $\alpha_{lj}\equiv 0$ at  $j<l$, and
\begin{equation}
  \lim_{r\rightarrow 0} \alpha_{lj} = a_l^{(j)} \left( \frac{r}{r_{m,n}}\right)^l
  (2l+1)\int\limits_{-1}^{1} P_l(x) x^l dx
  \label{eq:lim_alpha_lj}
\end{equation}
at $j\geq l$. Moreover, we can easily show that
$\alpha_{lj}\equiv 0$ when  $j>l$ and $j-l=1,3,5,...$ (i.e., when
the parities of
 $j$ and $l$ do not coincide), because the corresponding coefficients  $a_l^{(j)}$ equal zero.
Using the known values of the coefficients $a_l^{(j)}$ of the
$j$-th degree  Legendre polynomial and the integral in
\eqref{eq:lim_alpha_lj} (see ~\cite{prudnikov1986integrals,
bateman1953v1}), we obtain the expression
 \begin{equation*}
  \lim_{r \rightarrow 0} \alpha_{lj} = (2l+1)\cdot \left( \frac{r}{r_{m,n}}\right)^l \cdot
  \frac{2\cdot l!}{(2l+1)!!}\cdot
  \frac{1}{2^{j}} \cdot  (-1)^{(j-l)/2} \frac{j!}{(\frac{j-l}{2})!(\frac{j+l}{2})!}
  \frac{(j+l)!}{j! l!},
\end{equation*}
that holds true if $(j-l)$ is a nonnegative even number. In all
other cases, the coefficients $\alpha_{lj}$ are identically
zero.
The introduction of a new index $j'$ using the relation $j=l+2j'$
enables writing this expression in a more convenient form (the
prime in the new index is dropped):
 \begin{equation}
  \lim_{r \rightarrow 0} \alpha_{l, l+2j} = \left( \frac{r}{r_{m,n}}\right)^l
  \frac{1}{2^{l+2j-1}} \frac{(-1)^j}{(2l-1)!!}\frac{(2(l+j))!}{j!(l+j)!}.
  \label{eq:lim_alpha_lj_final}
\end{equation}

Substitution of \eqref{eq:lim_alpha_lj_final} into
\eqref{eq:psi_incident_mn}-\eqref{eq:Legendre_decompos} gives
\begin{multline}
  \Psi_{sc}^{(m,n)}=ik \sum\limits_{j=0}^{\infty} i^j F_j h_j(kr_{m,n})
  P_j\left(\frac{r}{r_{m,n}}\cos\theta\right) =
  ik\sum\limits_{l=0}^\infty P_l(\cos\theta) \left\{
    \sum\limits_{j=0}^\infty i^j F_j h_j(kr_{m,n})\frac{\alpha_{lj}}{2}
  \right\} = \\ =
ik \sum\limits_{l=0}^\infty
  i^l r^l P_l(\cos\theta)\frac{1}{2^l}\frac{1}{(2l-1)!!}\left\{
    \frac{1}{r_{m,n}^l}\sum\limits_{j=0}^\infty F_{l+2j}h_{l+2j}(kr_{m,n})\frac{1}{2^{2j}}
    \frac{(2(l+j))!}{j!(l+j)!} \right\},
  \label{eq:psi_incident_mn_decompos_proof}
\end{multline}
i.e., we have the expression identical to
\eqref{eq:psi_incident_mn_decompos1}.

For partial-wave expansion of $\Psi_{sc,0}$, we need to perform
summation of \eqref{eq:psi_incident_mn_decompos_proof} over all
$m,n$ but $(m,n)=(0,0)$. To do this, it suffices to find the
values of the sums $\sum\limits_{m,n}
h_{l+2j}(kr_{m,n})r_{m,n}^{-l}$, where $r_{m,n} = \sqrt{(am)^2 +
(bn)^2}$. As  stated in section 3, we are interested only in the
real parts of these sums. For convenience, let us introduce the
notation
 \begin{equation*}
  \Sigma_{lj}\equiv \Re \sum\limits_{(m,n)\neq (0,0)}  \frac{h_{l+2j}(kr_{m,n})}{r_{m,n}^{l}}.
 \end{equation*}
The first sum can be found using, for example, the Poisson
summation formula:
\begin{equation}
  \Sigma_{00} = \frac{2\pi}{k^2 a b} -1.
  \label{eq:sum_00}
\end{equation}
The following formula for the spherical Hankel functions is
helpful in seeking $\Sigma_{l0}$ at  $l>0$ \cite{Mors1953}:
\begin{equation}
  \frac{h_{l+1}(kr)}{r^{l+1}} = \frac{1}{k^{l+2}}\int\limits_0^k k^{l+2} \frac{h_l(kr)}{r^l}dk.
  \label{eq:hankel_int2}
\end{equation}
Using the mathematical induction method and
\eqref{eq:hankel_int2}, we can readily prove that
\begin{equation}
  \Sigma_{l0} = k^l \frac{1}{(2l-1)!!}\left\{  \frac{2\pi}{k^2 a b} - \frac{1}{2l+1} \right\}
  \label{eq:sum_l0}
\end{equation}
for all $l$.

To find the values of  $\Sigma_{lj}$ at $j>0$, let us make use of
the the recurrent relation between the Hankel functions in the
form \cite{Mors1953}
 \begin{equation}
  h_{l+2j}(kr) = \frac{2l+4j -1}{kr}h_{l+2j-1}(kr) - h_{l+2j-2}(kr).
  \label{eq:hankel_reccur}
\end{equation}
Substitution of $j=1,2,...$ into \eqref{eq:hankel_reccur} gives
the following values for the first few  sums
\begin{eqnarray*}
  \Sigma_{l1} = k^l\frac{2\pi}{k^2 a b}\frac{2}{(2l+1)!!},\\
  \Sigma_{l2} = k^l\frac{2\pi}{k^2 a b}\frac{8}{(2l+3)!!},\\
  \Sigma_{l3} = k^l\frac{2\pi}{k^2 a b}\frac{48}{(2l+5)!!},\\
  \cdots
\end{eqnarray*}

One can see that the general expression for  $\Sigma_{lj}$ at
$j>0$ will have the form
\begin{equation}
  \Sigma_{lj} = k^l\frac{2\pi}{k^2 a b} \frac{2^j j!}{(2l + 2j -1)!!}.
  \label{eq:sum_lj}
\end{equation}
The equality  \eqref{eq:sum_lj} can be rigorously proved using the
mathematical induction method.

Transform the double factorials in \eqref{eq:sum_l0} and
\eqref{eq:sum_lj} by formula $(2n+1)!!=\dfrac{(2n+1)!}{2^n \cdot
n!}$ and thus obtain the final expressions for  $\Sigma_{lj}$
\begin{eqnarray}
  \Sigma_{lj} = k^l\cdot 2^l \cdot \frac{2^{2j-1} (l+j-1)! j!}
  {(2l+2j-1)!}\cdot \frac{2\pi}{k^2 a b}, \;\;\; j>0,
  \label{eq:sum_lj_final1} \\
  \Sigma_{l0} = k^l \cdot 2^l \cdot \frac{2^{-1}(l-1)!}{(2l-1)!}\left\{
    -\frac{1}{2l+1} + \frac{2\pi}{k^2 a b} \right\} .
  \label{eq:sum_lj_final2}
\end{eqnarray}
Simple transformations of
\eqref{eq:psi_incident_mn_decompos_proof} and
\eqref{eq:sum_lj_final1}-\eqref{eq:sum_lj_final2}  give the sought
expansion for the wave $\Psi_{sc,0}$ when $r\rightarrow 0$
\begin{equation*}
   \Psi_{sc, 0} = \sum\limits_{(m,n)\neq(0,0)} \Psi_{sc}^{(m,n)} \approx
  ik \sum\limits_{l=0}^\infty i^l r^l k^l P_l(\cos\theta)\frac{1}{(2l-1)!!}
  \left\{ -\frac{1}{2l+1}F_l + \frac{2\pi}{k^2 a b}\sum\limits_{j=0}^\infty F_{l+2j} \right\}.
\end{equation*}

\end{document}